%% file: main.tex
\documentclass{article}
\usepackage{graphicx} % Required for inserting images
\usepackage{fullpage}
\usepackage{multirow}
\usepackage{float}
\usepackage{url}
\usepackage{amsfonts,amsmath,amssymb,amsthm}

\newtheorem{theorem}{Theorem}
\newtheorem{problem}{Problem}

\newtheorem{definition}{Definition}
\newtheorem{corollary}{Corollary}
\newtheorem{claim}{Claim}
\newcommand{\LEAN}{\includegraphics[height=.8em]{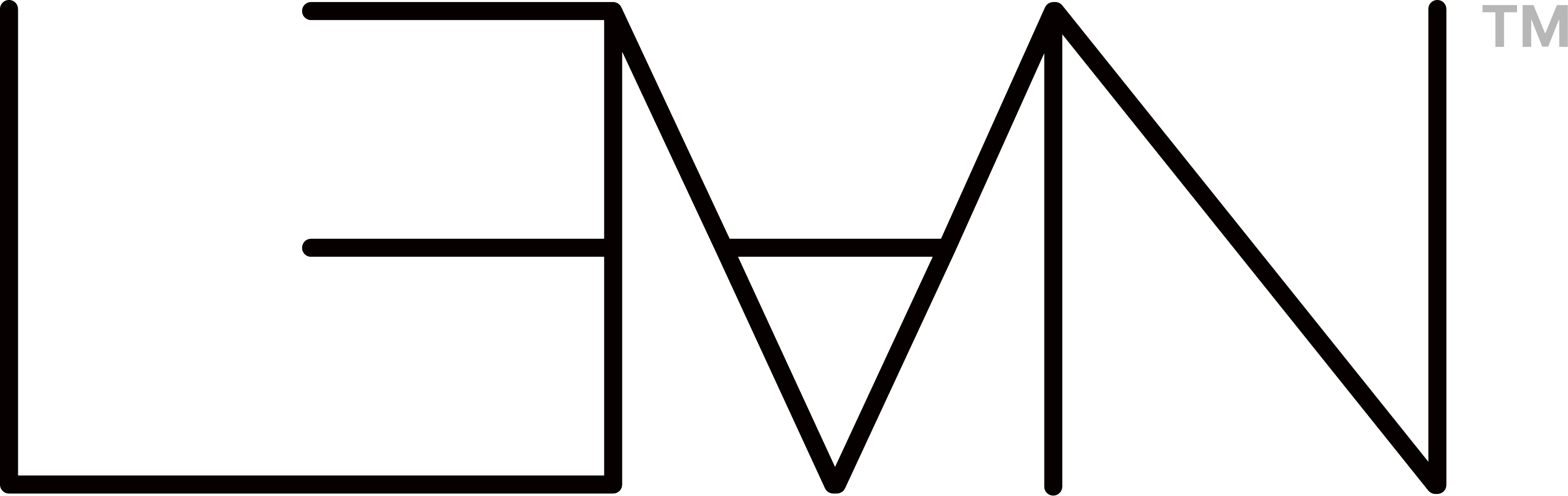}\hspace{-.1cm}}

\title{A Strongly Subquadratic Algorithm for Unique Assembly Verification in the Abstract Tile Assembly Model\thanks{This research was supported in part by National Science Foundation Grant CCF-2329918.}}
\author{Tim Wylie}
\date{timothy.wylie@utrgv.edu\\University of Texas Rio Grande Valley}

\begin{document}

\maketitle

\begin{abstract}
    We show, for the first time, a strongly subquadratic algorithm for Unique Assembly Verification (UAV) in the 2D Abstract Tile Assembly Model (aTAM) with any constant temperature. UAV in the 2D aTAM is solvable in $\widetilde{O}(n^{71/36})$. Further, in 3D, we give a logarithmic improvement for UAV to $O(n^2/\sqrt{\log(n+2)})$, and then show that even for assemblies with a constant height of two in 3D (barely 3D), no strongly subquadratic algorithm can exist under the Orthogonal Vectors Conjecture (OVC) by showing it is coOV-hard. Further, we show that UAV for barely 3D assemblies is P-complete and analyze the result with relation to Circuit-SETH. We show that for UAV, any 3D aTAM system can be simulated by a barely 3D aTAM system. This provides a clear conditional separation between the 2D and 3D problems (even with only two layers), and improves the time complexity for one of the most fundamental problems in algorithmic self-assembly in both 2D and 3D.
    %$O_\tau\!\left(n^{71/36}(1+\log(n+2))^{37/18}+(n+t)\log(n+t+2)\right)$
\end{abstract}

\input{intro}
\input{defs}

\input{log}

\input{ovhard}

\input{subquad}

%open questions
%is it ov-complete?
%k-uav?

%\section*{AI Usage Statement}
%The main results in Theorems \ref{thm:3d}, \ref{thm:2d}, and \ref{thm:ovc} were found using OpenAI 6 Astra. LEAN ... All writing ....

\bibliographystyle{plain}
\bibliography{uav}

\end{document}

%% file: intro.tex
\section{Introduction}\label{sec:intro}

The Abstract Tile-Assembly Model (aTAM) is possibly the best known model of self-assembly \cite{Winf98}. Although it is deceptively simple to define systems, it has its roots in Wang tiles \cite{Wang:1961:BSTJ}, and the model is Turing-complete \cite{Winf98}, Intrinsically Universal \cite{doty2012tile,hader2020impacts}, and has been studied extensively to for tasks such as the efficient assembly of shapes \cite{rothemund2000program}, and parallel computation \cite{BRUN2007,Keenan2016}. Many generalizations have also appeared, such as allowing for RNA tiles that can be deleted \cite{ABD2010SRS, DPS2010SAA}, multiple stages of growth \cite{Caballero:2021:ESA,CMS2018OSG,DFS2017NGA}, and even negative glues \cite{covert,Doty2013}. The aTAM is powerful because not only can the tile set store information, but work has also gone into using the seed \cite{need4seed}, or even the temperature \cite{Chalk:2018:ESA,highTemp}, for making systems more complex.

One of the most fundamental problems within the model is Unique Assembly Verification, which asks whether the tile system builds the desired assembly and nothing else \cite{Adleman:2002:STOC}. The initial algorithm, from \cite{Adleman:2002:STOC}, gives $O(n^2 + nt)$ where $n$ is the size of the final assembly and $t$ is the size of the tile set. This same algorithm also works for the 3D aTAM. Since this algorithm in 2002, no substantial progress has been achieved for an arbitrary temperature. For temperature-1 systems, UAV can be solved in $O(n \log t)$ \cite{Doty:2016:NC}.

\input{table}

\paragraph{AI Statement.}
The main results of this document were discovered through the use of OpenAI 6 Astra, and as part of the Algorithmic Self-Assembly Research Group (ASARG). The proofs are not currently included here, but the generated details and \LEAN\ verification code can be found publicly at \cite{Wylie:2026:GITHUB}. Full details will be included as the paper is written. This document was fully written by the authors, and they take responsibility for the claims herein.

\paragraph{Contributions.}
We present several new results showing improved running times for both 2D and 3D UAV, and a clear separation between 2D and 3D for UAV. See Table \ref{tab:results} for a comparison of the previous results for UAV in the aTAM along with the contributions from this work. For simplicity, we have assumed $t=O(n)$ in the table, where $t$ is the size of the tile set and $n$ the size of the assembly, however, the exact results are given in the theorems.

%soft-o notation ignores logs
%https://en.wikipedia.org/w/index.php?title=Big_O_notation&section=22#Extensions_to_the_Bachmann.E2.80.93Landau_notations

%% file: table.tex
\begin{table}[t]
    \centering
    \renewcommand{\arraystretch}{1.4}
    \begin{tabular}{|c|c | c|c | c |} \hline 
         \textbf{Model} & \textbf{Dim.} & \textbf{Temp.} & \textbf{Complexity} & \textbf{Reference} \\ \hline
        \multirow{7}{*}{aTAM} & \multirow{3}{*}{2} & 1 & $O(n \log n)$ & %Thm. 4.1 in 
        \cite{Doty:2016:NC} \\ \cline{3-5}
         &  & \multirow{2}{*}{$O(1)$} &$O(n^2)$ & %Thm. 3.2 in 
         \cite{Adleman:2002:STOC} \\ \cline{4-5}
         &  &  & $\widetilde{O}(n^{71/36})$ %or $O(n^{79/40})$ 
         %$O(n^{71/36}\Lambda^{37/18} + P)$
         & \textbf{Thm. \ref{thm:2d}} \\ \cline{2-5}
         %&  & $\tau$ & coNP-complete & Thm xx in \cite{} \\ \cline{2-5}
         & \multirow{4}{*}{3} & 1 & $O(n \log n)$ & %Thm x in 
         \cite{Doty:2016:NC} \\ \cline{3-5}
         &  & \multirow{4}{*}{$O(1)$} &$O(n^2)$ & %Thm. xx in 
         \cite{Adleman:2002:STOC} \\ \cline{4-5}
         &  &  & $O(n^2/\sqrt{\log(n+2)})$ & \textbf{Thm. \ref{thm:3d}} \\ \cline{4-5}
         &  &  & coOV-hard & \textbf{Thm. \ref{thm:ovc}} \\ \cline{4-5}
         &  &  & P-complete & \textbf{Thm. \ref{thm:pc}} \\ \hline
    \end{tabular}
    \caption{Known results for Unique Assembly Verification in the aTAM along with the new results. Here, $n$ is the number of tiles in the final assembly and $t$ is the size of the tile set. Note that $\widetilde{O}$ ignores logarithmic factors.
    %$\Lambda$ and $P$ are both logarithmic terms. 
    We have also assumed for simplicity that $t = O(n)$ since it simplifies the presentation. 
    }
    \label{tab:results}
\end{table}

%temp 1 in 3d

%% file: defs.tex
\section{Definitions}\label{sec:defs}

\paragraph{The Abstract Tile-Assembly Model.}
We provide a brief, high-level definition of the Abstract Tile-Assembly Model (aTAM) from \cite{GomezSchwellerWylie2025} and refer the reader to~\cite{rothemund2000program} for a formal definition. At a high level, the Abstract Tile-Assembly Model (aTAM) uses a set of \emph{tiles} capable of sticking together to construct shapes. These tiles are typically squares (2D) or cubes (3D) with \emph{glues} on each side where they may attach to one another. A glue is labeled to indicate its type, governing what other tiles it may bond with and the \emph{strength} of the bond. A tile with all of its labels is a \emph{tile type}. A \emph{tile set} contains all the tile types of the system. A single tile may attach at a location if the combined strength of the matching glues is greater than or equal to the \emph{temperature $\tau$}. An \emph{assembly} is a shape made up of one or more combined tiles. Construction is started around a designated \emph{seed} assembly \emph{S}. Any assembly capable of being made from the seed is called a \emph{producible} assembly. An assembly is \emph{terminal} if no more tiles can attach.  All producible assemblies have a path to grow into a terminal assembly. A terminal assembly is said to be \emph{uniquely produced} if it is the only terminal assembly that can be made by a tile system. A tile system is formally represented as an ordered triplet $\Gamma = (T, S, \tau)$ of the tile set, seed assembly, and temperature parameter, respectively. 

%\subsection{Problems}

\paragraph{Barely 3D.} For 3D aTAM systems that only have height-2 in the $Z$ dimension, the term \emph{barely 3D} has often been used in the self-assembly literature \cite{CookFuSchweller2011,DrakePatitzSummersTracy2025,FurcyMickaSummers2017,FurcySummersWithers2021}. We use a slightly stricter notion of a barely 3D aTAM system where any tile type may only appear in either the top or the bottom layer. That is, the tile set is a partition $T = \{T_u, T_l\}$ of all tiles where every tile can appear only in the $Z=0$ or $Z=1$ layer. Although not required, we specify every tile in the upper set $T_u$ has a null glue for their top face and every tile in $T_l$ has a null glue for their bottom face. Even if not explicitly stated, every reduction in this work is to this stricter notion.

\begin{problem}[Unique Assembly Verification]
Given an aTAM system $\Gamma = (T, S, \tau)$ and an assembly $A$, does $\Gamma$ uniquely build assembly $A$?
\end{problem}

%\paragraph{Orthogonal Vectors Conjecture} basaousantehuaohuoe \cite{Kane:2019:ITCS}

%\paragraph{Circuit-SETH}
%defn
%\cite{Lampis:2026:SODA}

%\paragraph{Circuit-SAT}

%\paragraph{Circuit Satisfying Pair} basaousantehuaohuoe 
%\cite{Abboud:2016:STOC}

%\paragraph{Circuit Value Problem (CVP)}

\subsection{Computational Problems and Hypotheses}

A Boolean circuit, as used here, is a finite directed acyclic graph with designated input vertices, \({\land,\lor,\neg}\) gates, and a designated output. We use 2-input AND and OR gates.  Let $v$ be the number of input variables and $s$ the number of gates and wires, excluding input vertices. %Circuits are represented explicitly, with gates listed in topological order.

%All algorithms and hypotheses below concern uniform deterministic algorithms in a fixed computational model, taken here to be the word-RAM with words of \(\Theta(\log L)\) bits for encoded input length (L). Randomized versions of the hypotheses are stated separately when needed. 

\begin{problem}[Circuit Satisfiability (CircuitSAT)]
Given a Boolean circuit $C:\{0,1\}^{v}\rightarrow\{0,1\}$, does $C$ have a satisfying assignment?
\end{problem}

If $C$ is a circuit with $v$ inputs that are partitioned into blocks of lengths \(r_A\) and \(r_B\), with \(v = r_A+r_B\). Then
$C\in\mathrm{CircuitSAT}$ is equivalent to 
$
\left(
C,\{0,1\}^{r_A},\{0,1\}^{r_B}
\right)\in\mathrm{CircuitSP}$.
This expresses CircuitSAT as a satisfying-pair search over the full Boolean domain. The input to CircuitSAT is \(C\), and the input CircuitSP is both lists. Thus, creating the full domain is an exponential-size transformation.
For the balanced partitions $r_A=\lfloor v/2\rfloor, r_B=\lceil v/2\rceil$, the lists have sizes $N_A=2^{\lfloor v/2\rfloor}, N_B=2^{\lceil v/2\rceil}$. 

There is another generalization of CircuitSAT that already has this concept. Weighted-Circuit-SAT with $k=2$ is equivalent to CircuitSP \cite{Abrahamson:1995:APAL}.

\begin{problem}[Circuit-Satisfying-Pair (CircuitSP) \cite{Abboud:2016:STOC}] 
Given a Boolean circuit $C:\{0,1\}^{r_A}\times\{0,1\}^{r_B}
\rightarrow \{0,1\}$, 
and two explicitly represented finite lists $A=(a_1,\ldots,a_{N_A})$ and $B=(b_1,\ldots,b_{N_B})$
where \(a_i\in\{0,1\}^{r_A}\) and \(b_j\in\{0,1\}^{r_B}\). Does some pair of listed strings satisfy the circuit $C$?
\end{problem}
%Thus, CircuitSP asks whether \(C\) has a satisfying assignment in the Cartesian product \(A\times B\). Its total input arity is \(v=r_A+r_B\). The balanced case has \(N_A=N_B=N\).

If we restrict CircuitSP such that \(N_A=N_B=1\), we get the Circuit Value Problem. CVP is P-complete under deterministic logarithmic-space many-one reductions.

\begin{problem}[Circuit Value Problem (CVP)]
Given a Boolean circuit $ C:\{0,1\}^{v}\rightarrow\{0,1\}$
and a fully specified assignment \(z\in\{0,1\}^{v}\), does $C(z) = 1$?
\end{problem}

%CVP is P-complete under deterministic logarithmic-space many-one reductions %, including when the circuit is supplied in topological order.

%Singleton-list specialization. For any partition \(z=(a,b)\) of the assignment into two blocks,
% \[
% (C,z)\in\mathrm{CVP}
% \quad\Longleftrightarrow\quad
% (C,{a},{b})\in\mathrm{CircuitSP}.
% \]

%Equivalently, substituting the supplied assignment into all input vertices gives a circuit with no free variables; that circuit is satisfiable exactly when the original CVP instance is a YES-instance.

\begin{problem}[Orthogonal Vectors (OV)]
Given two lists $A=(a_1,\ldots,a_{N_A}), B=(b_1,\ldots,b_{N_B})$ of vectors in \({0,1}^{d}\). Is there an orthogonal pair?
% Equivalently, the two vectors have no coordinate at which both are one.
% \[
% \bigwedge_{\ell=1}^{d}
% \neg(x_\ell\land y_\ell).
% \]

% This predicate has a bounded-fan-in circuit of size (O(d)), using a binary tree for the outer conjunction. By construction,
% \[
% (A,B)\in\mathrm{OV}
% \quad\Longleftrightarrow\quad
% (C_{\perp,d},A,B)\in\mathrm{CircuitSP}.
% \]
% OV is the restriction of CircuitSP to the prescribed orthogonality circuit \(C_{\perp,d}\).
\end{problem}

\begin{definition}[Orthogonal Vectors Conjecture (OVC) \cite{Kane:2019:ITCS}] 
For every constant \(\varepsilon\in(0,1)\), there exists a constant \(c\ge1\) such that no algorithm solves every balanced OV instance consisting of two lists of (N) vectors in dimension
$d=\lceil c\log N\rceil$ in time $O\left(N^{2-\varepsilon}\right)$.
Here \(N\ge2\). This is the logarithmic-dimensional formulation of OVC.

% The quantifier order is
% \[
% \forall\varepsilon\in(0,1);\exists c\ge1.
% \]
% In particular, the dimension constant (c) may depend on the proposed exponent saving \(\varepsilon\). 
\end{definition}

\begin{definition} [Circuit Strong Exponential Time Hypothesis (Circuit-SETH) \cite{Lampis:2026:SODA}] 
There do not exist constants \(\delta\in(0,1)\) and \(k\ge0\), and an algorithm deciding CircuitSAT for every circuit of size $s$ with $v$  variables, whose running time is $O\left( 2^{(1-\delta)v}(s+v+1)^k \right)$.

%Equivalently, for every fixed \(\eta\in(0,1)\), CircuitSAT has no algorithm running in $ (2-\eta)^v(s+v+1)^{O(1)}$ time. The exponential saving and the degree of the polynomial factor must be independent of the input circuit. This is the general-circuit formulation: circuit size and depth are unrestricted.
\end{definition}

%They can be constructed in time
%\[
%O!\left(s+v,2^{\lceil v/2\rceil}\right).
%\]
%If equal list lengths are required, the shorter list can be padded by repetitions. This is the split-and-list reduction.

\begin{corollary}
Assuming Circuit-SETH, there are no fixed constants \(\varepsilon\in(0,1)\) and \(k\ge0\), and an algorithm solving every balanced CircuitSP instance in time $O\left(N^{2-\varepsilon}(s+v+1)^k
\right)$.
\end{corollary}
% Proof. Apply the algorithm to the full-domain lists obtained by the balanced split above, with \(N=2^{\lceil v/2\rceil}\). Its running time, including construction of the lists, is
% \[
% O\left(
% 2^{(1-\varepsilon/2)v}(s+v+1)^{O(1)}
% \right),
% \]
% which contradicts Circuit-SETH. \(\square\)

% Complement problems

% The complementary problems used in uniqueness reductions are defined by reversing the existential answers:
% \[
% (C,A,B)\in\mathrm{coCircuitSP}
% \quad\Longleftrightarrow\quad
% \forall a\in A;\forall b\in B,\quad C(a,b)=0,
% \]
% and
% \[
% (A,B)\in\mathrm{coOV}
% \quad\Longleftrightarrow\quad
% \forall a\in A;\forall b\in B,\quad
% \sum_{\ell=1}^{d}a_\ell b_\ell>0.
% \]

% The orthogonality-predicate specialization also gives
% \[
% (A,B)\in\mathrm{coOV}
% \quad\Longleftrightarrow\quad
% (C_{\perp,d},A,B)\in\mathrm{coCircuitSP}.
% \]

%% file: log.tex
\section{Contributions}

% I would present the relationship as follows:
% 1. Circuit-Pair theorem: the central general construction.
% 2. coOV-hardness corollary: obtained by choosing the orthogonality predicate.
% 3. Circuit-SETH corollary: obtained by listing all half-assignments of an arbitrary circuit.
% 4. Seven-layer OV theorem: the sharper specialized construction, with \(O(N(d+1))\) size.

\subsection{A Logarithmic Improvement for 2D and 3D}\label{sec:log}
The first result is a logarithmic improvement to the running time for UAV in 3D. The simplification holds for 2D as well.  The improvement uses small square regions and tabulates every boundary input state.  The tabulation follows the general Four Russians principle \cite{ArlazarovEtAl1970,MasekPaterson1980} and shifting \cite{HochbaumMaass1985}.

% \begin{theorem}[2D/3D UAV]\label{thm:3d}
%      Given a 2D or 3D aTAM system with $\tau \geq 1$, UAV can be solved in time
%      \[
%     O \left ( \frac{n^2}{ \sqrt{\log(n + 2)}} + \Lambda \right )
%     \]
%     where $\Lambda = (n + t) \log(n + t + 2)$.
%     Exact randomized hashing of word-sized keys gives Las Vegas expected word-RAM time with $\Lambda = t$. 
%     Both use $O(n^{3/2} + t)$ cells, or $O(n^{1+\eta} + t)$ cells for any fixed rational $0 < \eta < 1$, with constants depending on $\eta$. 
%     When $t = O(n)$, both models give $O(n^{2}/\sqrt{\log(n + 2)})$.
%  \end{theorem}
\begin{theorem}[2D/3D UAV]\label{thm:3d}
     Given a 2D or 3D aTAM system with $\tau \geq 1$, UAV can be solved in time
     \[
    O \left ( \frac{n^2}{ \sqrt{\log(n + 2)}} + (n + t) \log(n + t + 2) \right )
    \]
     using $O(n^{3/2} + t)$ cells, or $O(n^{1+\eta} + t)$ cells for any fixed rational $0 < \eta < 1$, with constants depending on $\eta$. 
    When $t = O(n)$, this gives $O(n^{2}/\sqrt{\log(n + 2)})$.
 \end{theorem}
 
% \begin{theorem}[2D/3D UAV]\label{thm:3d}
%     Given a 2D or 3D aTAM system with $\tau \geq 1$ s.t. $t=O(n)$, UAV can be solved in
%     \[
% O \left ( \frac{n^{2}}{\sqrt{\log(n + 2)}} \right )
% \]
% time using $O(n^{3/2})$ space, or $O(n^{1+\eta})$ cells for any fixed rational $0 < \eta < 1$. %, with constants depending on $\eta$. 
% \end{theorem}

%% file: ovhard.tex
\subsection{3D UAV Separation}\label{sec:ovhard}
We showed a logarithmic improvement over quadratic time exists for 3D UAV in Theorem \ref{thm:3d}, which we complement now with lower bounds and some analysis. First, we show that 3D UAV is coOV-hard, which by the Orthogonal Vectors Conjecture, implies that no strongly subquadratic algorithm exists (even for a constant height assembly of two). We then give a more general reduction from Circuit-Satisfying-Pair to nonUAV, which can be viewed as a generalization of OV (still with a constant height of two). Applying this construction to all pairs of half-assignments yields a Circuit-SAT reduction, which also rules out a truly subquadratic algorithm under general Circuit-SETH. Finally, we show that 3D UAV is P-complete using the Circuit Value Problem.

Although coOV-hardness is technically implied by the other results in this section, we mention it here since coOV-hardness is more established and it demonstrates another aspect of the flexibility of 3D tile systems.

\begin{theorem}[3D UAV is coOV-hard]\label{thm:ovc}
    Given a 3D aTAM system with $\tau \geq 1$, Unique Assembly Verification is coOV-hard even with assemblies of height two.
\end{theorem}

The following theorem is proven with a log-space reduction from CVP to UAV of a height-2 assembly with a singleton seed at temperature 2.

\begin{theorem}[3D UAV is P-complete]\label{thm:pc}
    Given a 3D aTAM system with $\tau \geq 1$, Unique Assembly Verification is P-complete even with assemblies of height two and a singleton seed.
\end{theorem}

Although this next result does not prove anything related to runtime, we provide it here since the construction and its existence imply that all the power from 3D aTAM systems is due to having two layers of tiles. The proof is not provided, but the basic technique is similar to the CVP reduction. Basically the individual height layers can be viewed as being placed as slabs adjacent to each other in the 2D plane. However, in building each layer, tile placement is actually controlled by OR and AND gates, which place tiles based on whether the original system could have placed those tiles. 

We forego a formal definition of simulation, but for this result, we mean that all producible assembly paths have an equivalent path in the simulation. Due to communication lag in the simulation, this is a many-to-one onto mapping from the simulating system to the original. Thus, we only guarantee that it is a parsimonious reduction for UAV. We are listing it as a claim rather than a theorem since the exact simulation definition is not provided here.

\begin{claim}[Slab Simulation]\label{thm:slab}
    Every 3D UAV instance can be transformed, with polynomial overhead, into an equivalent UAV instance whose simulator is confined to two layers, operates at temperature two, and has a singleton seed.
\end{claim}

\subsubsection{Circuit-SETH}

Although not a typical hardness class, we use the CircuitSP problem to prove quadratic lower bounds under Circut-SETH.

\begin{theorem}[3D UAV is CircuitSP-hard]\label{thm:csphard}
    Given a 3D aTAM system with $\tau \geq 1$, Unique Assembly Verification is coCircuitSP-hard even with assemblies of height two.
\end{theorem}

\begin{corollary}
Assuming Circuit-SETH, for every fixed \(\varepsilon>0\), there is no algorithm deciding 3D UAV in $O\left((n+t)^{2-\varepsilon}\right)$,
where \(n\) is the size of the target assembly and \(t\) is the number of tile types. This conditional lower bound holds even when \(t=O(n)\) and the height of the assembly is limited to two.    
\end{corollary}

%% file: subquad.tex
\subsection{A Subquadratic Algorithm for 2D}\label{sec:subquad}
Here we state the results achieved by exploiting planar properties of the tileset. By preprocessing small group interactions, we can decompose the assembly into smaller boundary areas and deal with regional interactions and boundary conditions. By carefully selecting the group size, we can achieve the following.

\begin{theorem}[2D UAV]\label{thm:2d}
    Given a 2D aTAM system with $\tau \geq 1$, UAV can be solved in 
    \[
    O(n^{71/36}\Lambda^{37/18} + P), \mbox{ s.t.} \quad \Lambda = 1 + \log(n+2),\quad P = (n+t) \log(n+t+2)
    \]
    time
    %with $\Lambda = 1 + \log(n+2)$ and $P = (n+t) \log(n+t+2)$ 
    using $O(n^{19/12}\Lambda^{11/6}+n+t)$ space.
    If $t=O(n)$, then 2D UAV is solvable in $O(n^{79/40})=O(n^{1.975})$ time using $O(n^{19/12}\Lambda^{11/6} )$ space.
\end{theorem}

\begin{figure}[t]
    \centering
    \includegraphics[width=1.\linewidth]{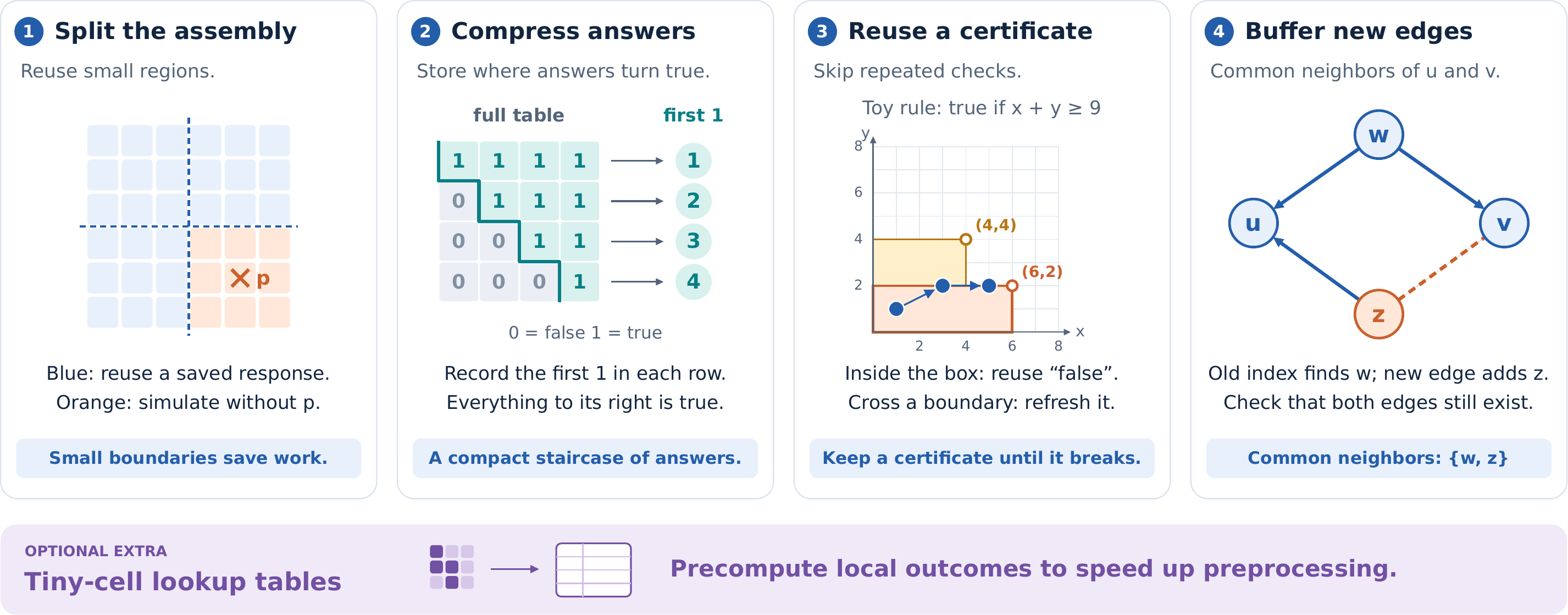}
    \caption{An AI image overview of the main techniques being used. }
    \label{fig:2duav}
\end{figure}

Although we are not covering the details, the full \LEAN\ verification is available \cite{Wylie:2026:GITHUB}. The algorithm uses planar graph separators, a peeling method to remove tiles/vertices dependent on preexisting tiles, knowledge of ordering constraints for placement, a simplification of boundary states, geometric range searches, sparse contact graphs, threshold-frontier compression, event-driven monitoring, and preprocessing groups of tile interactions. Figure \ref{fig:2duav} gives a brief AI overview of the most important techniques being used to achieve the subquadratic runtime.

% \begin{proof}
%     The proof is through a full \LEAN\ verification. The Lean theorem is....

%     the code is at \cite{Wylie:2026:GITHUB}
% \end{proof}